\documentclass[12pt]{article}
\usepackage{epsf,epsfig,a4,latexsym}
\usepackage{amsmath}
\usepackage{amssymb}
\usepackage{amsfonts}

\newcommand{\threequart}{\mbox{\small $\frac{3}{4}$}}    
\newcommand{\msbar}{\mbox{\tiny $\overline{MS}$}}        
\newcommand{\qq}{\mbox{\tiny $q\overline{q}$}}           
\newcommand{\lat}{\mbox{\tiny $L\!A\!T$}}                
\newcommand{\plaq}{\Box}                                 

\def\lsim{\mathrel{\rlap{\lower4pt\hbox{\hskip1pt$\sim$}}
    \raise1pt\hbox{$<$}}}                
\def\gsim{\mathrel{\rlap{\lower4pt\hbox{\hskip1pt$\sim$}}
    \raise1pt\hbox{$>$}}}                
\begin{document}

\title{
\vspace{-3cm}
\flushleft{\normalsize DESY 05-028} \\
\vspace{-0.4cm}
{\normalsize Edinburgh 2005/02} \\
\vspace{-0.4cm}
{\normalsize LTH 647} \\
\vspace{-0.4cm}
{\normalsize LU-ITP 2005/012} \\
\vspace{-0.4cm}
{\normalsize February 2005} \\
\vspace{0.4cm}
\centering{\Large \bf A Determination of the Lambda Parameter from Full
  Lattice QCD}}

\author{\large M. G\"ockeler$^{1,2}$, R. Horsley$^3$,
               A.~C. Irving$^4$, D. Pleiter$^5$, \\
               P.~E.~L. Rakow$^4$, G. Schierholz$^{5,6}$
               and H. St\"uben$^7$ \\[1em]
         -- QCDSF-UKQCD Collaboration -- \\[1em]
        \small
           $^1$ Institut f\"ur Theoretische Physik,
                Universit\"at Leipzig, \\[-0.5em]
        \small
                D-04109 Leipzig, Germany \\[0.25em]
        \small
           $^2$ Institut f\"ur Theoretische Physik,
                Universit\"at Regensburg, \\[-0.5em]
        \small
                D-93040 Regensburg, Germany \\[0.25em]
        \small
           $^3$ School of Physics, University of Edinburgh, \\[-0.5em]
        \small 
                Edinburgh EH9 3JZ, UK \\[0.25em]
        \small
           $^4$ Department of Mathematical Sciences, University of Liverpool,
                                                            \\[-0.5em]
        \small
                Liverpool L69 3BX, UK \\[0.25em]
        \small
           $^5$ John von Neumann-Institut f\"ur Computing NIC, \\[-0.5em]
        \small
                Deutsches Elektronen-Synchrotron DESY, \\[-0.5em]
        \small
                D-15738 Zeuthen, Germany \\[0.25em]
        \small
           $^6$ Deutsches Elektronen-Synchrotron DESY, \\[-0.5em] 
        \small
                D-22603 Hamburg, Germany \\[0.25em]
        \small
           $^7$ Konrad-Zuse-Zentrum f\"ur Informationstechnik Berlin,
                                                       \\[-0.5em]
        \small
                D-14195 Berlin, Germany}

\date{ }

\maketitle


\begin{abstract}
We present a determination of the QCD parameter $\Lambda$ in the quenched
approximation ($n_f=0$) and for two flavours ($n_f=2$) of light dynamical
quarks. The calculations are performed on the lattice using $O(a)$ improved
Wilson fermions and include taking the continuum limit. We find
$\Lambda^{\msbar}_{n_f=0} = 259(1)(20) \, \mbox{MeV}$ and 
$\Lambda^{\msbar}_{n_f=2} =261(17)(26) \, \mbox{MeV}$, using $r_0=0.467 \,
\mbox{fm}$ to set the scale. Extrapolating our results
to five flavours, we obtain for the running coupling constant at the mass of
the $Z$ boson $\alpha_s^{\msbar}(m_Z)=0.112(1)(2)$.

\end{abstract}

\clearpage


\section{Introduction}


The parameter $\Lambda$ is one of the fundamental quantities of
QCD. It sets the scale for the running coupling constant $\alpha_s(\mu)$,
and it is the only parameter of the theory in the chiral limit.
Usually $\Lambda$ is defined by writing $\alpha_s(\mu)$ as an expansion
in inverse powers of $\ln(\mu^2/\Lambda^2)$. For such a relationship
to remain valid for all values of $\mu$, $\Lambda$ must change as
flavour thresholds are crossed: $\Lambda \rightarrow \Lambda_{n_f}$,
where $n_f$ indicates the effective number of light (with respect to the
scale $\mu$) quarks. 

A lattice calculation of $\Lambda$ requires an accurate determination of a
reference scale, the introduction of an appropriate non-perturbatively defined
coupling, which can be computed accurately on the lattice over a sufficiently
wide range of energies, as well as a reliable extrapolation to the chiral and
continuum limits. Finally, and equally importantly, one needs to know the
relation of the coupling to $\alpha_s^{\msbar}$, the quantity of final
interest, accurately to a few percent. This programme has been achieved for
the pure gauge theory~\cite{capitani98a,booth01a}. In full QCD calculations
with Wilson fermions the amount of lattice data was barely enough to enable a 
reliable chiral and continuum extrapolation~\cite{booth01a,booth01b}. 
Recent calculations with staggered fermions cover a wider range of
lattice spacings and quark masses~\cite{davies03}. However, staggered fermions
are not without their own problems. 

We determine $\Lambda$ in the $\overline{MS}$ scheme from the force parameter
$r_0$ \cite{sommer93a} and the `boosted' coupling $g_\plaq$. The latter is
obtained from the average plaquette. The advantage of this method is that both
quantities are known to high precision. As in our previous
work~\cite{booth01a,booth01b}, we shall use here non-perturbatively $O(a)$
improved Wilson (clover) fermions. Definitions of the action are
standard (see, for example, Appendix~D of \cite{gockeler04c}). The lattice
calculations will be done for $n_f=2$ flavours of dynamical quarks. In
addition, we will update our quenched results. 

Since our first attempt~\cite{booth01a,booth01b} the amount of 
lattice data with dynamical quarks has greatly increased~\cite{gockeler04a}. 
That is to say, at our previous couplings $\beta=5.20$, $5.25$ and $5.29$
we have increased the statistics and done additional simulations at smaller
quark masses. Furthermore, we have generated dynamical gauge field
configurations at $\beta=5.40$ for three different quark masses. At each
$\beta$ value we now have data at three to four quark masses at our
disposal, and the smallest lattice spacing that we have reached in our
simulations is $a\approx 0.07\,\mbox{fm}$. This allows us to improve on, and
disentangle, the chiral and continuum extrapolations.  In the quenched case
the force parameter $r_0/a$ is now known up to $\beta=6.92$~\cite{necco01a}.

The paper is organised as follows. In section~\ref{introduction} we present 
a general discussion about the $\beta$ function, including
Pad{\'e} approximations, and the running coupling constant. Also given are
results in the $\overline{MS}$ 
scheme. In section~\ref{section_theory}  we set up the lattice formalism and
discuss 
what coefficients are known. Various possibilities for converting to the
$\overline{MS}$ scheme are given, which will indicate the magnitude
of systematic errors. In section~\ref{results} results are given for
$r_0\Lambda^{\msbar}$ for both quenched ($n_f=0$) and unquenched $n_f=2$ 
fermions. These results are then extrapolated to $n_f=3$ flavours of dynamical
quarks in section~\ref{force_scales}. This is done by matching the static
force at the scale $r_0$. In section~\ref{phenomenology} we convert our
results to physical units and, after matching $\alpha_s$ to $n_f=5$ flavours,
compare them with other lattice determinations and to the experimental
values. Finally, in section~\ref{conclusions} we give our conclusions.


\section{The QCD Coupling and the $\beta$ Function}
\label{introduction}


The `running' of the QCD coupling constant as the scale changes is
controlled by the $\beta$ function,
\begin{equation}
   {\partial g_{\cal S}(M) \over \partial \log M }
                           = \beta^{\cal S} g_{\cal S}(M) 
\label{defbeta}
\end{equation}
with
\begin{equation}
   \beta^{\cal S} \left(g_{\cal S}\right)
       = - b_0g_{\cal S}^3 - b_1g_{\cal S}^5
         - b_2^{\cal S}g_{\cal S}^7 
         - b_3^{\cal S}g_{\cal S}^9 - \ldots \,,
\label{defbfun}
\end{equation}
renormalisation having introduced a scale $M$ together with a scheme
$\cal S$. The first two coefficients are scheme independent and are
given for the $SU(3)$ colour gauge group as
\begin{equation}
   b_0 = {1\over (4\pi)^2}
           \left( 11 - {2\over 3}n_f \right) \,, \qquad
   b_1 = {1\over (4\pi)^4}
           \left( 102 - {38 \over 3} n_f \right) \,.
\label{b0+b1}
\end{equation}
Integrating eq.~(\ref{defbeta}) gives
\begin{equation}
   {\Lambda^{\cal S} \over M} = F^{\cal S}(g_{\cal S}(M)) \,,
\end{equation}
with
\begin{equation}
   F^{\cal S}(g_{\cal S})
     = \exp{\left( - {1\over 2b_0 g_{\cal S}^2}\right)} 
         \left(b_0 g_{\cal S}^2 \right)^{- {b_1\over 2b_0^2}}
         \exp{\left\{ - \int_0^{g_{\cal S}} \! d\xi
         \left[ {1 \over \beta^{\cal S}(\xi)} +
                {1\over b_0 \xi^3} - {b_1\over b_0^2\xi} \right]\right\} } \,,
\label{bfunsol}
\end{equation}
where $\Lambda^{\cal S}$, the integration constant, is the fundamental
scheme dependent QCD parameter. The integral in eq.~(\ref{bfunsol})
may be performed numerically or to low orders analytically. For
example, to 3 loops we have
\begin{equation}
   {\Lambda^{\cal S} \over M}
     = \exp{\left( - {1\over 2b_0 g_{\cal S}^2}\right)} 
         \left(b_0 g_{\cal S}^2 \right)^{- {b_1\over 2b_0^2}}
   \left( 1 + {A^{\cal S}\over 2b_0} g_{\cal S}^2 \right)^{-p^{\cal S}_A}
   \left( 1 + {B^{\cal S}\over 2b_0} g_{\cal S}^2 \right)^{-p^{\cal S}_B} \,,
\label{three_loop_exact}
\end{equation}
where
\begin{equation}
\begin{split}
   A^{\cal S} &= b_1 + \sqrt{b_1^2 - 4b_0b_2^{\cal S}} \,, \\
   B^{\cal S} &= b_1 - \sqrt{b_1^2 - 4b_0b_2^{\cal S}} \,,
\end{split}
\end{equation}
and
\begin{equation}
\begin{split}
   p^{\cal S}_A
       &= - {b_1 \over 4b_0^2} -  {b_1^2 - 2b_0b_2^{\cal S} \over 4b_0^2 
                                     \sqrt{b_1^2 - 4b_0b_2^{\cal S}} } \,,
                                          \\
   p^{\cal S}_B
       &= - {b_1 \over 4b_0^2} +  {b_1^2 - 2b_0b_2^{\cal S} \over 4b_0^2 
                                     \sqrt{b_1^2 - 4b_0b_2^{\cal S}} } \,.
\end{split}
\end{equation}

\begin{figure}[t]
\begin{center}
   \epsfxsize=10.00cm \epsfbox{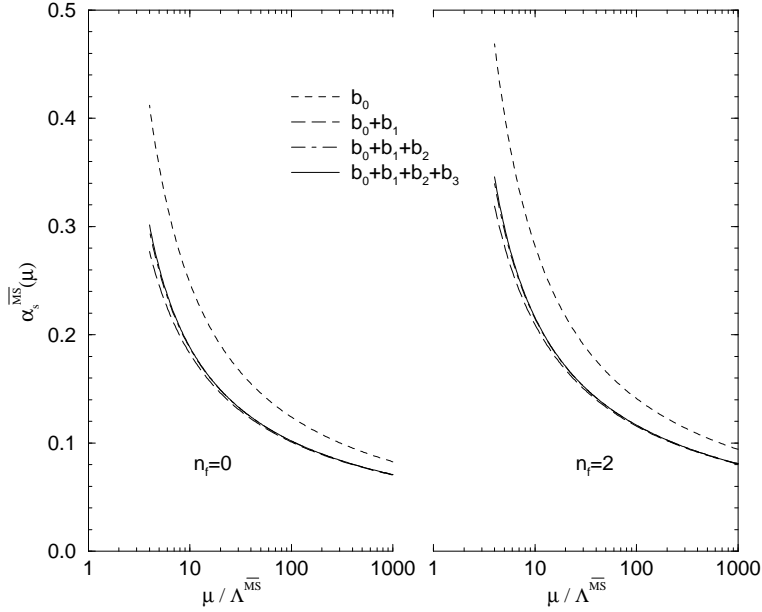}
\end{center}
   \caption{$\alpha_s^{\msbar}(\mu)$ versus
            $\mu/\Lambda^{\msbar}$ for $n_f=0$ (left picture)
            and $n_f=2$ (right picture), using successively
            more and more coefficients of the $\beta$ function.}
   \label{fig_alpha_nf0+nf2_mu}
\end{figure}

Results are usually given in the $\overline{MS}$ scheme, with the scale
$M$ being replaced by $\mu$, and thus
\begin{equation}
   {\Lambda^{\msbar} \over \mu} = F^{\msbar}(g_{\msbar}(\mu)) \,.
\label{lambda_msbar}
\end{equation}
In this scheme the next two $\beta$ function coefficients are 
known~\cite{tarasov80a,larin93a,vanritbergen97a}:
\begin{equation}
\begin{split}
   b_2^{\msbar}
        &= {1\over (4\pi)^6}
            \left( {2857\over 2} - {5033\over 18}n_f 
                    + {325\over 54}n_f^2 \right)  \,,  \\
   b_3^{\msbar}
       &= {1\over (4\pi)^8}
           \left[ {149753\over 6} + 3564\,\zeta_3 \right.
         - \left({1078361\over 162} + {6508\over 27}\zeta_3\right)n_f  \\
       &   \left. \hspace*{3.23cm}
            + \left({50065\over 162}+ {6472\over 81}\zeta_3\right)n_f^2
            + {1093\over 729}n_f^3 \right] \,.
\end{split}
\label{beta_msbar}
\end{equation}
The running coupling  $\alpha^{\msbar}_s(\mu) \equiv g^2_{\msbar}(\mu)/4\pi$
is plotted in Fig.~\ref{fig_alpha_nf0+nf2_mu} for $n_f=0$, $2$
by solving eq.~(\ref{bfunsol}) numerically, using only the 
first coefficient (1-loop), the first and second coefficients (2-loop)
etc.\ of the $\beta$ function. The figure shows an apparently rapidly
convergent series (cf the 3- to 4-loop result), certainly in the range we will
be interested in, $\mu/\Lambda^{\msbar} \sim 20$. The main difference between
the $n_f=0$ and $n_f=2$ results is that $\alpha_s^{\msbar}|_{n_f=2}$
rises more steeply as a function of $\mu/\Lambda^{\msbar}$,
as $b_0|_{n_f=2} < b_0|_{n_f=0}$.

\begin{figure}[htb]
\begin{center}
   \epsfxsize=10.00cm \epsfbox{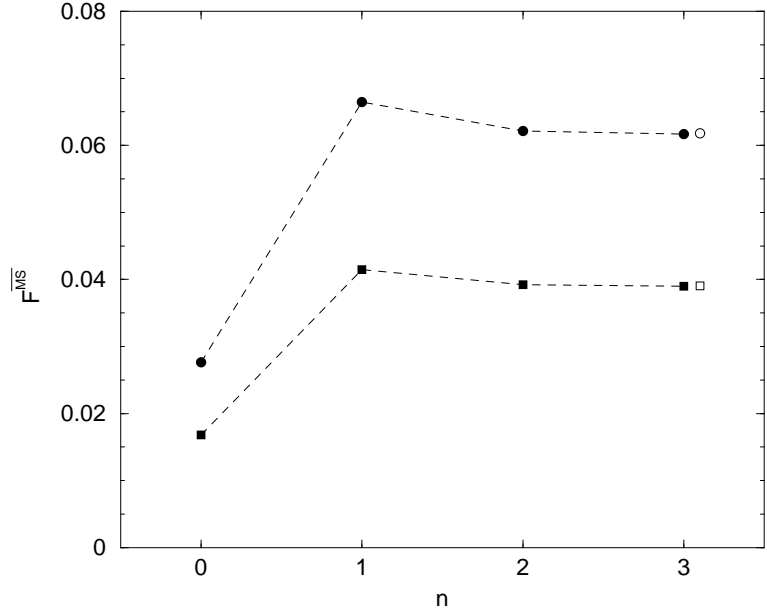}
\end{center}
   \caption{$F^{\msbar}(g_{\msbar})$ for $g_{\msbar}^2 = 2$ versus
            $\beta$ function coefficient number $n$. The $n_f=0$ 
            values are filled circles, while the $n_f=2$ values
            are filled squares. The $[1/1]$ Pad{\'e} approximations
            are given as open symbols.}
\label{fig_F_from_g2MSbar=2_nf0+nf2}
\end{figure}

A knowledge of the $\beta$ function to 4 loops is the exception
rather than the rule. In many schemes it is known only to 3 loops.
To improve the convergence of the $\beta$ function, we may attempt
to use a Pad{\'e} approximation by writing eq.~(\ref{defbfun}) as
\begin{equation}
   \beta^{\cal S}_{[1/1]}(g_{\cal S}) = 
      - { b_0g_{\cal S}^3 + 
          \left( b_1 - {b_0b_2^{\cal S}\over b_1} \right) g_{\cal S}^5 \over
          1 - {b_2^{\cal S}\over b_1} g_{\cal S}^2 } \,,
\label{beta_11pade}
\end{equation}
which on expanding is arranged to give the first three coefficients 
of eq.~(\ref{defbfun}) and estimates the next coefficient
$b_3^{\cal S}$ as
\begin{equation}
   b_3^{\cal S} \approx { (b_2^{\cal S})^2 \over b_1} \,.
\label{b3approx}
\end{equation}
It is again possible to give an analytic result for $F^{\cal S}$
using $\beta^{\cal S}_{[1/1]}$. We find
\begin{equation}
   {\Lambda^{\cal S} \over M}
     = \exp{\left( - {1\over 2b_0 g_{\cal S}^2}\right)}
         \left[ b_0 g_{\cal S}^2 \over 1 + \left( {b_1\over b_0} - 
                                                  {b_2^{\cal S} \over b_1}
                                           \right) g_{\cal S}^2
   \right]^{-{ b_1 \over 2b_0^2}} \,.
\label{beta_11pade_exact}
\end{equation}
At least for the $\overline{MS}$ scheme this appears to work reasonably
well. Equation~(\ref{b3approx}) gives $b_3^{\msbar} \approx 3.22\times 10^{-5}$
and $1.67\times 10^{-5}$ for quenched and unquenched fermions,
respectively, to be compared with the true values from eq.~(\ref{beta_msbar})
of $4.70\times 10^{-5}$ and $2.73\times 10^{-5}$. In \cite{booth01b}
we have shown a figure of the various Pad{\'e} approximations to the
$\beta$ function. In Fig.~\ref{fig_F_from_g2MSbar=2_nf0+nf2}
we show the value of $F^{\msbar}(g_{\msbar})$ at $g_{\msbar}^2 = 2$ versus
the $\beta$ function coefficient number for both quenched and 
unquenched fermions. Also shown are the results using the $[1/1]$ Pad{\'e}
approximations. It is seen that these numbers lie extremely close to
the 4-loop $\beta$ function results.
As Pad{\'e} approximations give some estimation
of the effect of higher order $\beta$ function coefficients, we shall
thus prefer these later in our determination of the $\Lambda$ parameter.

\section{Lattice Methods}
\label{section_theory}


On the lattice we also have a coupling constant $g_0(a)$ and corresponding
$\beta$ function with coefficients $b_i^{\lat}$ and parameter
$\Lambda^{\lat}$, where
\begin{equation}
   a \Lambda^{\lat}  = F^{\lat}(g_0(a)) \,.
\label{lambda_lat}
\end{equation}
To evaluate $F^{\lat}$, we need to know the $b_i^{\lat}$s. They can
be found by expanding $g_{\msbar}$ as a power series in $g_0$ as
\begin{equation}
\begin{split}
   {1 \over g_{\msbar}^2(\mu)}
     &= {1 \over g_0^2(a)} + 2b_0 \ln a\mu - t^{\lat}_1
          + (2b_1 \ln a\mu - t^{\lat}_2) g_0^2(a)  \\
     &+ [-2b_0b_1 \ln^2 a\mu + 2(b_2^{\msbar} + b_1t_1^{\lat})
     \ln a\mu 
          - t_3^{\lat} ]\, g^4_0(a)
          + \ldots \,.
\end{split}
\label{gmsglat}  
\end{equation}
To have consistency between eqs.~(\ref{lambda_msbar}) and (\ref{lambda_lat})
we need
\begin{equation}
   t_1^{\lat} = 2 b_0 \ln {\Lambda^{\msbar}\over \Lambda^{\lat}} \,,
\end{equation}
and
\begin{equation}
\begin{split}
   b_2^{\lat} &= b_2^{\msbar} + b_1t^{\lat}_1 - b_0t^{\lat}_2     \,,
                                         \\
   b_3^{\lat} &= b_3^{\msbar} + 2b_2^{\msbar}t_1^{\lat} 
                   + b_1 (t_1^{\lat})^2 - 2b_0 t^{\lat}_3          \,,
\end{split}
\label{bi_lat}
\end{equation}
where $b_i^{\lat}$ are the lattice $\beta$ function coefficients,
as in eq.~(\ref{defbfun}).
So the transformation between the two schemes is given by the
$t_i^{\lat}$ (which define the transformation), and the renormalisation
group dictates how the scale running occurs (in this case the
$\ln a\mu$ terms). A knowledge of (the 1-loop) $t_1^{\lat}$ determines
the relationship between the $\Lambda$ parameters in the two schemes,
while also knowing (the 2-loop) $t_2^{\lat}$ means that the 3-loop
$\beta$ function coefficient $b_2^{\lat}$ can be found.

At present, what we know is
\cite{luscher98a,luscher98b,alles98a,sint96a,marcantonio00a,
bode01a,booth01a}
\begin{equation}
\begin{split}
   t_1^{\lat} &= 0.4682013 - n_f [0.0066960 - 0.0050467\,c_{sw}  +
   0.0298435\,c_{sw}^2\\ 
   &+ am_q ( -0.0272837 + 0.0223503\,c_{sw}  - 0.0070667\,c_{sw}^2 ) +
   O((am_q))^2] \,, 
   \\[0.5em]
   t_2^{\lat} &=  0.0556675 - n_f [ 0.002600+0.000155\,c_{sw} -
   0.012834\,c_{sw}^2\\ 
       &- 0.000474\,c_{sw}^3 - 0.000104\,c_{sw}^4  + O(am_q)]  \,. 
\end{split}
\label{known_ti}
\end{equation}
Here $t_1^{\lat}$ has been calculated including the $am_q$ terms ($m_q$
being the bare quark mass),
while $t_2^{\lat}$ is known only for $am_q = 0$, and $t_3^{\lat}$
is unknown, which means that from eq.~(\ref{bi_lat})
$b_2^{\lat}$ is known but not $b_3^{\lat}$.
For general $c_{sw}$ the connection between $g_{\msbar}^2$ and $g_0^2$
is only defined up to terms of $O(a)$, but on the improvement 
trajectory $c_{sw} = 1 + O(g_0^2)$ it is possible to arrange it 
to be $O(a^2)$ if the $am_q$ terms are included in the $t_i^{\lat}$s.

Thus, the conversion from the lattice
coupling to the $\overline{MS}$ coupling (eqs.~(\ref{gmsglat}) and
(\ref{known_ti})) can also be written with 
mass independent $t_i^{\lat}$s, if we redefine $g_0^2$ by replacing
it by $\tilde{g}_0^2$, where
\begin{equation}
   \tilde{g}_0^2 = g_0^2 ( 1 + b_g am_q )  \,,
      \qquad b_g = b_g^{(0)} n_f g_0^2 + O(g_0^4) \,.
\end{equation}
So, putting $c_{sw} = 1 + O(g_0^2)$ into eqs.~(\ref{gmsglat}) and
(\ref{known_ti})  
means that $t^{\lat}_1$ is replaced by $t^{\lat}_1 - n_f\, am_qb_g^{(0)}$,
which gives $b_g^{(0)} = 0.01200$. This value agrees with the number reported
in \cite{luscher96a}.

Thus, in this mass independent scheme (ie a scheme where the renormalisation
conditions are imposed for zero quark mass) there appears to be little
difference in extrapolating to the chiral limit using constant
$\beta = 6/g_0^2$, rather than constant $\tilde{\beta} = 6/\tilde{g_0}^2$.
So, rather than using eq.~(\ref{known_ti}) at finite $am_q$, we shall
first extrapolate our plaquette and $r_0/a$ data to the chiral limit
and then determine $\Lambda^{\msbar}$.
Before attempting this, we shall discuss some improvements to help 
improve the convergence of the power series (\ref{gmsglat}).

As it is well known that lattice perturbative expansions are poorly
convergent, we have used a `boosted' coupling constant
\begin{equation}
   g_{\plaq}^2 \equiv { g_0^2(a) \over u_0^4 } 
\end{equation}
to help the series (\ref{gmsglat}), or equivalently (\ref{defbfun})
for $\beta^{\lat}(g_0)$, converge faster. Here
$P \equiv u_0^4 = \langle \mbox{Tr}U^{\plaq}\rangle / 3$ is the average
plaquette. In perturbation theory we write
\begin{equation}
   {1\over g_{\plaq}^2} = {1\over g_0^2} - p_1 - p_2g_0^2 + O(g_0^4) 
\end{equation}
with \cite{bali02a,athenodorou04a}
\begin{equation}
\begin{split}
   p_1 &= {1\over 3}  \,,  \\
   p_2 &= 0.0339110 - n_f (0.001846 - 0.0000539\, c_{sw} + 0.001590\,c_{sw}^2) 
\end{split} \label{check}
\end{equation}
for massless clover fermions.

To improve the convergence of the series further,
we re-express it in terms of the tadpole improved coefficient
\begin{equation}
   c_{sw}^{\plaq} = c_{sw} u_0^3 \,.
\label{csw_original}
\end{equation}
Changing $t_i^{\lat}$ to $t_i^{\plaq}$ first replaces
$t_i^{\lat}$ by $t_i^{\lat} - p_i$, and secondly using $c_{sw}^{\plaq}$
simply replaces every $c_{sw}$ by $c_{sw}^{\plaq}$ in $t_1^{\lat}$,
but the change in $t_2^{\plaq}$ is more complicated as the coefficients
of $c^{\plaq}_{sw}$ change in $t_2^{\plaq}$.

This gives for 
$t_i^{\plaq} \equiv t_i^{\plaq}(c_{sw}^{\plaq})$ in the chiral limit
\begin{equation}
\begin{split}
  t_1^{\plaq} 
       &= 0.1348680 - n_f [0.0066960 - 0.0050467\,c^{\plaq}_{sw}
                              + 0.0298435\,(c^{\plaq}_{sw})^2]  \\
   t_2^{\plaq} &= 0.0217565 - n_f [ 0.000753+0.001053\,c^{\plaq}_{sw}
                                  - 0.000498\,(c^{\plaq}_{sw})^2   \\
       & - 0.00047\,4(c^{\plaq}_{sw})^3 
           - 0.000104\,(c^{\plaq}_{sw})^4]  \,.
\end{split}
\label{known_ti_plaq}
\end{equation}
As we have here a 2-loop result, we can see how well tadpole
improvement improves the series convergence. The coefficient of
$n_f$ in $t_2^{\plaq}$ is considerably smaller
than the corresponding coefficient in $t_2^{\lat}$.
For example, using the values at $\beta = 5.40$ given in the
next section, we find that the magnitude of the coefficient is
reduced by two orders of magnitude (from $\sim -0.0438$ to $\sim 0.0003$).

What this tadpole improvement represents is taking a path from $g^2 = 0$
to $g^2 = g_{\plaq}^2$, keeping $c_{sw}^{\plaq}$ fixed. Later we shall
consider other trajectories from $0$ to $g_{\plaq}^2$. If we had all orders
of the theory, the result would depend only on the end point.
But with a finite series the trajectory will matter. This will help us
estimate systematic errors from unknown higher order terms.

Thus, in conclusion we have
\begin{equation}
   a \Lambda^{\plaq} = F^{\plaq}(g_{\plaq}(a)) \,, 
\label{lambda_plaq}
\end{equation}
\begin{equation}
   {\Lambda^{\msbar} \over \mu} = F^{\msbar}(g_{\msbar}(\mu)) \,,
\label{lambda_msbarr}
\end{equation}
together with the conversion formula
\begin{equation}
   {1 \over g_{\msbar}^2(\mu)}
     = {1 \over g_{\plaq}^2(a)} + 2b_0 \ln a\mu - t^{\plaq}_1
          + (2b_1 \ln a\mu - t^{\plaq}_2) g_{\plaq}^2(a)
          + \ldots 
\label{gmsbarglat_full}
\end{equation}
with
\begin{equation}
   t_1^{\plaq} = 2 b_0 \ln {\Lambda^{\msbar}\over \Lambda^{\plaq}} 
\end{equation}
and
\begin{equation}
   b_2^{\plaq} = b_2^{\msbar} + b_1t^{\plaq}_1 - b_0t^{\plaq}_2   \,.
\label{bi_plaq}
\end{equation}

We shall now discuss various strategies to determine $\Lambda^{\msbar}$.


\subsection{Method I}


This method was used in our previous
papers~\cite{booth01a,booth01b,gockeler04b}, with the difference 
that now we first extrapolate to the chiral limit. For each $\beta$ value
we first compute $t_i^{\plaq}$ from eq.~(\ref{known_ti_plaq}).
Then from eq.~(\ref{gmsbarglat_full}) we convert $g_{\plaq}$ to 
$g_{\msbar}$ at some appropriate scale $\mu_*$, and using the force
scale $r_0$, we calculate $r_0 \Lambda^{\msbar}$
from eq.~(\ref{lambda_msbarr}):
\begin{equation}
   r_0 \Lambda^{\msbar} = r_0 \mu_* F^{\msbar}( g_{\msbar}(\mu_*))  \,.
\label{r0lambda}
\end{equation}
Finally, we extrapolate to the continuum limit, $a \to 0$.
Note that $t_i^{\plaq}$ will depend on the coupling because
$c_{sw}^{\plaq}$ does.

We must determine the scale $\mu_*$. A good choice to help
eq.~(\ref{gmsbarglat_full}) converge rapidly is to take
the $O(1)$ coefficient to vanish, which is achieved
by choosing \cite{luscher98b}
\begin{equation}
   \mu_* = {1\over a} \exp \left( {t_1^{\plaq} \over 2b_0} \right)  \,.
\end{equation}
Thus, we used
\begin{equation}
   {1 \over g_{\msbar}^2(\mu_*)}
     = {1 \over g_{\plaq}^2(a)}
          + \left( {b_1 \over b_0} t^{\plaq}_1 - t^{\plaq}_2 \right)
              g_{\plaq}^2(a) + O(g_{\plaq}^4) 
\label{gmsgplaq}
\end{equation}
to find $g_{\msbar}^2(\mu_*)$, which was then substituted into
eq.~(\ref{r0lambda}).


\subsection{Method II}


Alternatively, we can first determine
$b_2^{\plaq}$ from eq.~(\ref{bi_plaq}) and then determine
$r_0\Lambda^{\plaq}$ via eq.~(\ref{lambda_plaq}).
After computing this, we convert to $r_0\Lambda^{\msbar}$ using
\begin{equation}
   r_0\Lambda^{\msbar} = r_0\Lambda^{\plaq} 
                           \exp{ \left( t_1^{\plaq} \over 2 b_0 \right) } \,,
\label{lam_plaq_to_lam_msbar}
\end{equation}
and then take the continuum limit.
Again, note that $b_2^{\plaq}$ will depend on the
coupling, because $c_{sw}^{\plaq}$ does.

This method is equivalent to choosing a scale $\mu_=$, as in method I,
such that $g_{\msbar}(\mu_=) = g_{\plaq}(a)$. In this case {\it all}
the coefficient terms of eq.~(\ref{gmsbarglat_full}) vanish.
The scale that achieves this is
\begin{equation}
   \mu_= = {1\over a} \exp \left( {t_1^{\plaq} \over 2b_0} \right)
               { F^{\plaq}(g_{\plaq}(a)) \over F^{\msbar}(g_{\plaq}(a))} \,.
\label{mu_equal}
\end{equation}
Indeed, substituting $\mu_=$ into eq.~(\ref{lambda_msbarr})
then gives eq.~(\ref{lam_plaq_to_lam_msbar}) again.
The scale $\mu_=$ is close to $\mu_*$, as can be seen by expanding
eq.~(\ref{mu_equal}) to 3 loops. From eq.~(\ref{three_loop_exact})
we have
\begin{equation}
\begin{split}
   \mu_= &= {1\over a} \exp \left( {t_1^{\plaq} \over 2b_0} \right)
   { \left( 1 + {A^{\plaq}\over 2b_0} g_{\plaq}^2 \right)^{-p^{\plaq}_A}
     \over
     \left( 1 + {A^{\msbar}\over 2b_0} g_{\plaq}^2 \right)^{-p^{\msbar}_A}
   }
   { \left( 1 + {B^{\plaq}\over 2b_0} g_{\plaq}^2 \right)^{-p^{\plaq}_B}
     \over
     \left( 1 + {B^{\msbar}\over 2b_0} g_{\plaq}^2 \right)^{-p^{\msbar}_B}
   } 
                                            \\
         &= \mu_*
             \left( 1  - { b_1t_1^{\plaq} - b_0t_2^{\plaq} \over 2 b_0^2 }
                           g_{\plaq}^2 + \ldots \right) 
                                              \\
         &> \mu_* \,,
\end{split}
\end{equation}
for the couplings used here.


\subsection{Method III}


Another possibility, and theoretically the most sound, is to vary
$c_{sw}^{\plaq}$ along the improvement path as $g_{\plaq}^2$ increases. 
This will give genuinely constant $\beta$ function coefficients (ie
independent of the coupling).
As the 1-loop expansion for $c_{sw}^{\plaq}$ is known along this path,
\begin{equation}
   c^{\plaq}_{sw} = 1 + c_0^{\plaq} g_{\plaq}^2 + \ldots \,, 
\label{csw_plaq}
\end{equation}
with $c^{\plaq}_0 = c_0 - \threequart p_1$ and $c_0 = 0.2659(1)$
\cite{luscher96b}, then expanding eq.~(\ref{known_ti_plaq}) gives
\begin{equation}
   b_2^{\plaq} = b_2^{\msbar} + b_1 t_1^{\plaq}\big|_{c_{sw}^{\plaq}=1} - b_0
   t_2^{\plaq}\big|_{c_{sw}^{\plaq}=1} 
                 - b_0 c_0^{\plaq} \left. 
                   {\partial t_1^{\plaq} \over \partial c_{sw}^{\plaq}}
                      \right|_{c_{sw}^{\plaq}=1} 
               = -0.0008241 \,.
\label{b2_plaq_RG}
\end{equation}

This result may also be derived from eq.~(\ref{gmsbarglat_full})
by first setting $a = \mu^{-1}$ (for simplicity) and then taking
$\mu \partial / \partial\mu$ of this equation. This leads to
\begin{equation}
   - {2 \over g_{\msbar}^3} \beta^{\msbar}(g_{\msbar})
      = \left[ - {2 \over g_{\plaq}^3} 
                - {\partial t_1^{\plaq} \over \partial c_{sw}^{\plaq}}
                  {\partial c_{sw}^{\plaq} \over \partial g_{\plaq}}
                - 2t_2^{\plaq} g_{\plaq} + O(g_{\plaq}^3)
        \right]  \beta^{\plaq}(g_{\plaq}) \,,
\end{equation}
which upon expanding out also gives eq.~(\ref{b2_plaq_RG}).

So, having determined $b_2^{\plaq}$ in eq.~(\ref{bi_plaq}), the method is
as for method II: first determine $r_0\Lambda^{\plaq}$ 
using eq.~(\ref{lambda_plaq}) and then
convert to $r_0\Lambda^{\msbar}$ using eq.~(\ref{lam_plaq_to_lam_msbar}).

\subsection{Methods IIP and IIIP}

To further improve our calculations, and to reduce the systematic error, we
consider here the effect of Pad{\'e} improving the $\beta$ function, as given
in eqs.~(\ref{beta_11pade}) and (\ref{beta_11pade_exact}). We restrict
ourselves to methods II and III, and we call the Pad{\'e} improved results IIP
and IIIP, respectively.

\begin{table}[t]
   \begin{center}
      \begin{tabular}{||c|c|c|c|c|c||}
         \hline
         \hline & & & & & \\[-0.75em]
\multicolumn{1}{||c|}{$\beta$}                                & 
\multicolumn{1}{c|}{$r_0/a$}                                  & 
\multicolumn{1}{c|}{$P$}                                      &
\multicolumn{1}{c|}{$r_0\Lambda^{\msbar}$ I}                  &
\multicolumn{1}{c|}{$r_0\Lambda^{\msbar}$ II}                 &
\multicolumn{1}{c||}{$r_0\Lambda^{\msbar}$ IIP}               \\
         \hline
5.70  & 2.922(09) & 0.549195(25) & {\it 0.4888(15)} 
                                 & {\it 0.4950(15)} & {\it 0.4888(15)} \\
5.80  & 3.673(05) & 0.567651(21) & {\it 0.5142(07)}
                                 & {\it 0.5200(07)} & {\it 0.5140(07)} \\
5.95  & 4.898(12) & 0.588006(20) & 0.5461(13) & 0.5514(14) & 0.5457(13) \\
6.00  & 5.368(33)
                  & 0.593679(08) & {\it 0.5579(34)}
                                 & {\it 0.5631(35)} & {\it 0.5575(34)} \\
6.07  & 6.033(17) & 0.601099(18) & 0.5696(16) & 0.5746(16) & 0.5692(16) \\
6.20  & 7.380(26) & 0.613633(02) & 0.5861(21) & 0.5907(21) & 0.5855(21) \\
6.40  & 9.740(50) & 0.630633(04) & 0.5976(31) & 0.6018(31) & 0.5970(31) \\
6.57  & 12.18(10) & 0.643524(15) & 0.6029(48) & 0.6067(48) & 0.6022(48) \\
6.69  & 14.20(12) & 0.651936(15) & 0.6055(50) & 0.6091(51) & 0.6049(50) \\
6.81  & 16.54(12) & 0.659877(13) & 0.6080(46) & 0.6113(46) & 0.6073(46) \\
6.92  & 19.13(15) & 0.666721(12) & {\it 0.6145(47)}
                                 & {\it 0.6177(47)} & {\it 0.6139(47)} \\
         \hline
$\infty$ & $\infty$ & 1          & 0.6152(21) & 0.6189(21) & 0.6145(20) \\
         \hline
         \hline
      \end{tabular}
   \end{center}
\caption{The quenched $r_0\Lambda^{\msbar}$ values for methods I, II and IIP
         (ie using the Pad{\'e} improved $\beta$ function
         $\beta^{\plaq}_{[1/1]}$) together with the force parameter
         $r_0/a$ \cite{necco01a} (the number at $\beta=6.0$ is from the
         interpolation formula given there) and the plaquette $P$.
         The continuum extrapolated values together with the statistical
         errors are given in the bottom row. Numbers in {\it italics} are
         not used in the fits.}
\label{table_nf0}
\end{table}

\begin{figure}[!b]
\vspace*{-0.5cm}   
\begin{center}
   \epsfxsize=10.00cm \epsfbox{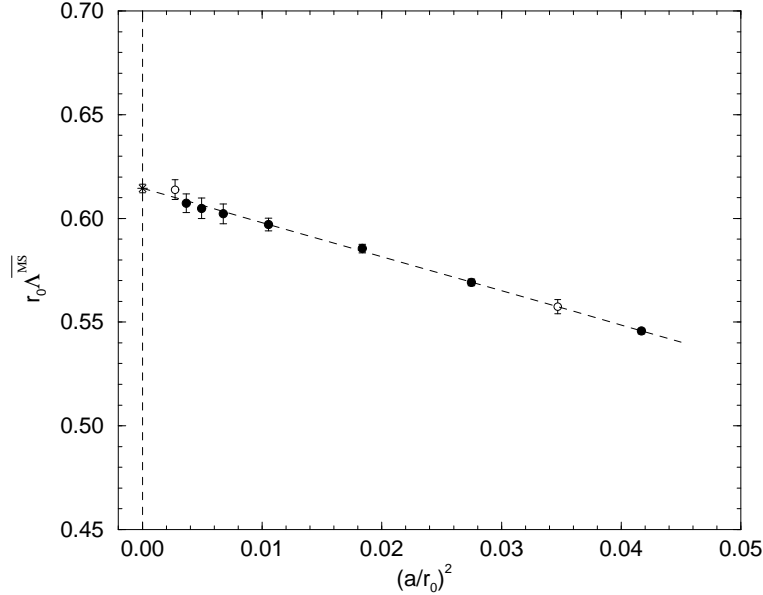}
\end{center}
   \caption{The quenched $r_0\Lambda^{\msbar}$ points versus $(a/r_0)^2$, 
            together with a linear extrapolation to the continuum
            limit for method IIP. The filled circles are
            used for the extrapolation. The star represents the
            extrapolated value.}
   \label{fig_r0lamMSbar_lbfun_nf0_IIP_8pts}
\end{figure}


\section{Results}
\label{results}


\subsection{Quenched Results}


In the quenched case ($n_f=0$) we do not have any of the additional
chiral limit extrapolation complications alluded to in the previous
section, or a $c_{sw}$ term. This means that there is no difference
between method II and method III, so the procedure is straightforward.
In Table~\ref{table_nf0} we give the parameters used.
For $r_0$ we use, for consistency, exclusively the values given
in \cite{necco01a}, which includes previous results from
\cite{guagnelli98a}. The one exception is $\beta = 6.0$,
where we have used the interpolation formula \cite{necco01a} for $r_0/a$.
Our plaquette values are determined at their given $\beta$ values.

In Table~\ref{table_nf0} we also give the results for
$r_0\Lambda^{\msbar}$ from methods I, II and IIP.
We first see that the results
for $r_0\Lambda^{\msbar}$ are almost indistinguishable between
methods I, II and IIP. Method IIP lies just below method I
(and indeed is almost identical with it).

We now consider the continuum limit of our results.
In Fig.~\ref{fig_r0lamMSbar_lbfun_nf0_IIP_8pts}
we plot the results for $r_0\Lambda^{\msbar}$ against $(a/r_0)^2$
for method IIP. The differences between the results of the various methods are
small. As one expects that Pad{\'e} improvement gives a better answer, we shall
concentrate on IIP.
The smallest $a$ value is not included in the fit, as it appears
to deviate a little, but including it would not have changed the
extrapolated value much.
We also have not included $\beta = 6.0$ in the fit, as $r_0/a$ is only
known from an interpolation formula. But as can be seen from the
figure, including it has no effect on the result.
Also, the two coarsest $a$ values have not
been included in the fit, as they show significant non-linear effects
in $a^2$. These two points are not shown in the plot, as they lie far to the
right. Figure~\ref{fig_r0lamMSbar_lbfun_nf0_IIP_8pts} clearly shows
a linear extrapolation over a wide range of lattice spacings,
$a^{-1} \sim 2$ -- $6.5\,\mbox{GeV}$, giving a value for method IIP of
\begin{equation}
 r_0\Lambda^{\msbar}_0 \equiv \left. r_0\Lambda^{\msbar}\right|_{n_f=0} =
 0.614(2)(5)  \,. 
\label{res0}
\end{equation}
Here the first error is statistical, and the second systematic error
is estimated by the spread in the results between methods I, II and IIP.
That the systematic error is small is an indication of the
convergence of results from the different methods. The result (\ref{res0})
agrees with our earlier value~\cite{booth01a}.


\subsection{Unquenched $n_f=2$ Results}


\begin{table}[!thp]
   \begin{center}
      \begin{tabular}{||c|l|c|l|l|l|l||}
         \hline
         \hline & & & & & & \\[-0.75em]
\multicolumn{1}{||c|}{$\beta$}                                & 
\multicolumn{1}{c|}{$\kappa$}                                 & 
\multicolumn{1}{c|}{$V$}                                      & 
\multicolumn{1}{c|}{$c_{sw}$}                                 & 
\multicolumn{1}{c|}{$r_0/a$}                                  & 
\multicolumn{1}{c|}{$P$}                                      & 
\multicolumn{1}{c||}{Group}                                   \\
         \hline
 5.20 & 0.1342    & $16^3\times 32$ & 2.0171 
      & 4.077(70) & 0.528994(58)    & QCDSF  \\
 5.20 & 0.1350    & $16^3\times 32$ & 2.0171
      & 4.754(45) & 0.533670(40)    & UKQCD  \\
 5.20 & 0.1355 & $16^3\times 32$ & 2.0171
      & 5.041(53) & 0.536250(30) & UKQCD  \\
 5.20 & 0.13565   & $16^3\times 32$ & 2.0171
      & 5.250(75) & 0.537070(100)   & UKQCD  \\
 5.20 & 0.1358 & $16^3\times 32$ & 2.0171
      & 5.320(95) & 0.537670(30)    & UKQCD  \\
         \hline
 5.25 & 0.1346    & $16^3\times 32$   & 1.9603       
      & 4.737(50) & 0.538770(41)    & QCDSF  \\
 5.25 & 0.1352    & $16^3\times 32$ & 1.9603       
      & 5.138(55) & 0.541150(30)    & UKQCD  \\
 5.25 & 0.13575   & $24^3\times 48$ & 1.9603       
      & 5.532(40) & 0.543135(15)   & QCDSF  \\
         \hline 
 5.29 & 0.1340    & $16^3\times 32$ & 1.9192
      & 4.813(82) & 0.542400(50)    & UKQCD  \\
 5.29 & 0.1350    & $16^3\times 32$ & 1.9192
      & 5.227(75) & 0.545520(29)    & QCDSF  \\
 5.29 & 0.1355    & $24^3\times 48$ & 1.9192
      & 5.566(64) & 0.547094(23)    & QCDSF  \\
 5.29 & 0.1359    & $24^3\times 48$ & 1.9192
      & 5.880(100)& 0.548286(57)    & QCDSF \\
         \hline
 5.40 & 0.1350    & $24^3\times 48$ & 1.8228
      & 6.092(67) & 0.559000(19)    & QCDSF  \\
 5.40 & 0.1356    & $24^3\times 48$ & 1.8228
      & 6.381(53) & 0.560246(10)    & QCDSF  \\
 5.40 & 0.1361    & $24^3\times 48$ & 1.8228
      & 6.714(64) & 0.561281(08)    & QCDSF  \\
         \hline
         \hline
      \end{tabular}
   \end{center}
\caption{The unquenched $\beta$, $\kappa$ and $c_{sw}$ values and the volume V,
         together with the measured force parameter $r_0/a$ and
         plaquette $P$. The collaboration that generated
         the configurations is given in the last column. The results for
         $\beta=5.29$, $\kappa=0.1359$ are preliminary. We have reanalysed our
         $r_0/a$ values, taking autocorrelations properly into account, which
         gave larger error bars than previously reported~\cite{alikhan04a}.}
\label{table_nf2}
\vspace*{1cm}
   \begin{center}
      \begin{tabular}{||c|l|l|l||}
         \hline
         \hline & & & \\[-0.75em]
\multicolumn{1}{||c|}{$\beta$}                                & 
\multicolumn{1}{c|}{$\kappa_c$}                               & 
\multicolumn{1}{c|}{$r_0/a$}                                  & 
\multicolumn{1}{c||}{$P$}                                     \\
         \hline
 5.20 & 0.136008(15) &  5.455(96)   & 0.538608(49)   \\
 5.25 & 0.136250(07) &  5.885(79)   & 0.544780(89)   \\
 5.29 & 0.136410(09) &  6.254(99)   & 0.549877(109)  \\
 5.40 & 0.136690(22) &  7.390(26)   & 0.562499(46)   \\
         \hline
         \hline
      \end{tabular}
   \end{center}
\caption{The critical values for $\kappa$ (ie $\kappa_c$)
         and the chiral limit values for $r_0/a$ and $P$ 
         for the four $\beta$ values used here.}
\label{table_critical}
\end{table}

\begin{table}[t]  \vspace*{-0.5cm}
   \begin{center}
      \begin{tabular}{||c|l|l|l|l|l||}
         \hline
         \hline  & & & & & \\[-0.75em]
\multicolumn{1}{||c|}{$\beta$}                                & 
\multicolumn{1}{l|}{$r_0\Lambda^{\msbar}$ I}                  & 
\multicolumn{1}{l|}{$r_0\Lambda^{\msbar}$ II}                 & 
\multicolumn{1}{l|}{$r_0\Lambda^{\msbar}$ IIP}                & 
\multicolumn{1}{l|}{$r_0\Lambda^{\msbar}$ III}                &
\multicolumn{1}{l||}{$r_0\Lambda^{\msbar}$ IIIP}              \\
         \hline
 5.20 & 0.5183(91)  & 0.5304(94)  & 0.4913(87)  & 0.6459(114) & 0.6173(109)\\
 5.25 & 0.5210(71)  & 0.5415(73)  & 0.5040(68)  & 0.6450(87)  & 0.6174(83) \\
 5.29 & 0.5372(85)  & 0.5482(87)  & 0.5120(81)  & 0.6433(102) & 0.6165(98) \\
 5.40 & 0.5577(198) & 0.5676(201) & 0.5343(189) & 0.6431(228) & 0.6182(219)\\
         \hline
$\infty$& 0.6012(346)& 0.6085(352)& 0.5819(329) & 0.6376(412) & 0.6170(395)\\
         \hline
         \hline
      \end{tabular}
   \end{center}
\caption{The values for $r_0\Lambda^{\msbar}$ for methods I, II, IIP,
         III, IIIP described in section~\ref{section_theory} for the four
         $\beta$ values used here.}
\label{table_r0_lam_dyn}
\end{table}

\begin{figure}[p]
   \begin{center}
\hspace*{-0.75cm}   \epsfxsize=10.00cm \epsfbox{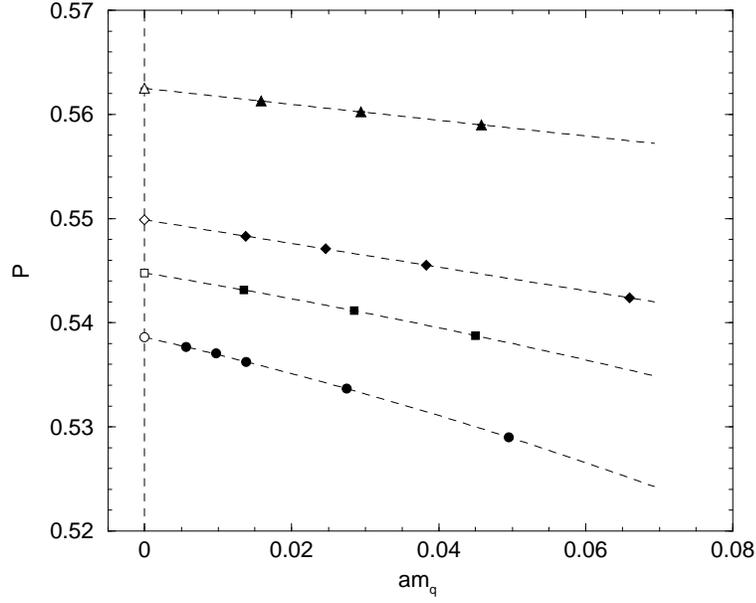}
   \end{center}
   \caption{The plaquette $P$ (filled symbols) plotted against the bare
            quark mass $am_q$ for $\beta=5.20$ (lower curve) until
            $\beta = 5.40$ (upper curve). The fits use
            eq.~(\ref{plaq_fit_eq}), giving the extrapolated values
            in the chiral limit (open symbols).}
\label{fig_amq_plaq_nf2_quad}
\end{figure}

\begin{figure}[p]
   \begin{center}
   \epsfxsize=10.00cm 
      \epsfbox{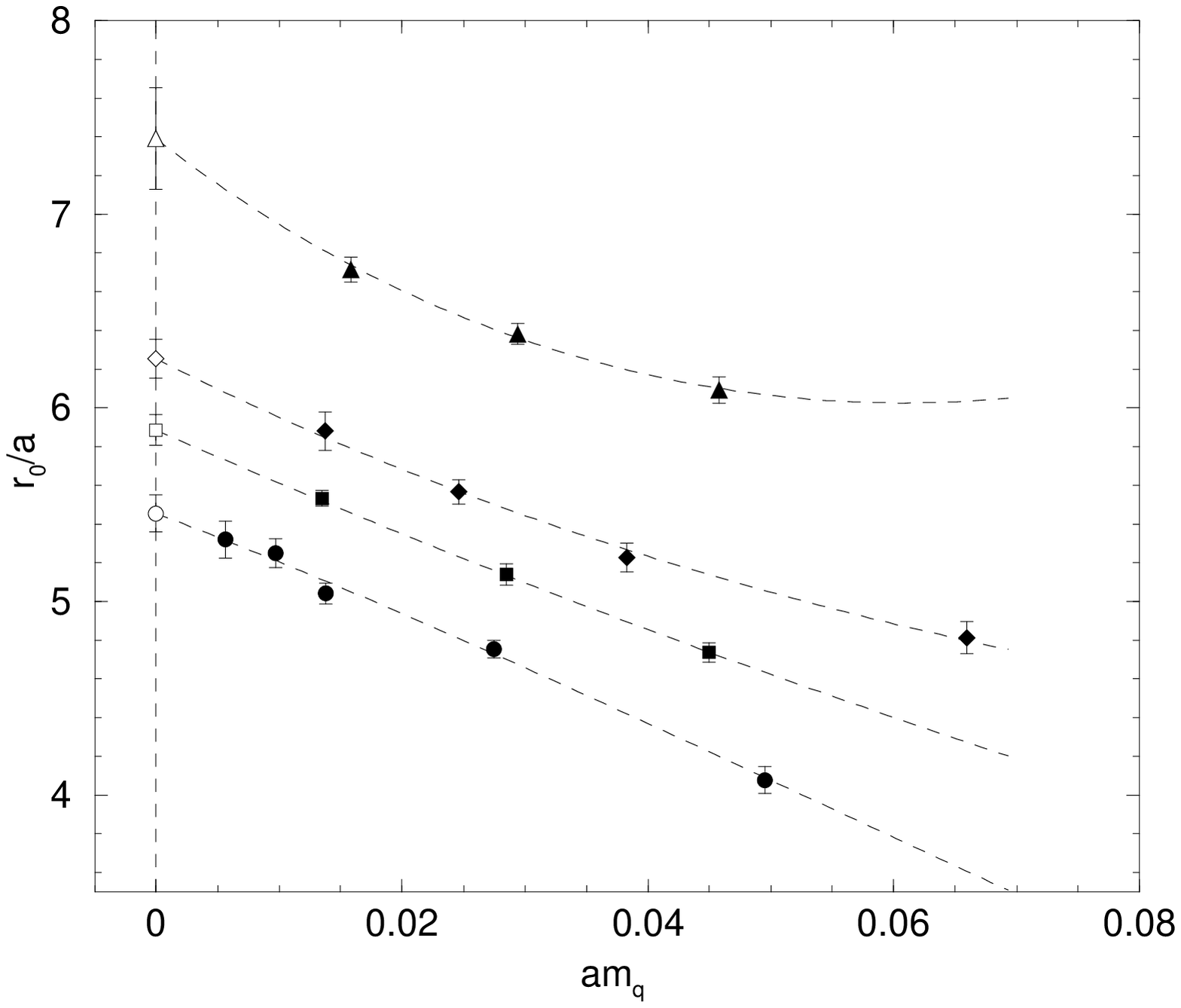}
   \end{center}
   \caption{The force parameter $r_0/a$ plotted against $am_q$. The same
      notation as in Fig.~\ref{fig_amq_plaq_nf2_quad} is used.}
\label{fig_amq_r0oa_nf2_joint8para_15pts_041221}
\end{figure}

We now turn to unquenched $n_f=2$ fermions. In Table~\ref{table_nf2}
we show the $\beta$, $\kappa$ and $c_{sw}$ parameters used in the simulations,
together with the measured $r_0/a$ and plaquette $P$ values.
As discussed in section~\ref{section_theory}, we shall first determine 
$r_0\Lambda^{\msbar}$ in the chiral limit and then perform the
continuum extrapolation. We must thus first find the zero quark mass
results from Table~\ref{table_nf2}. We shall make a chiral
extrapolation in $am_q$, defined here by
\begin{equation}
    am_q = {1\over 2} \left( {1\over\kappa} - {1\over\kappa_c} \right) \,.
\end{equation}
We estimate $\kappa_c$ from partially quenched pion data. The results have been
given in \cite{gockeler04a} and are tabulated in the second column of
Table~\ref{table_critical}. 

In Fig.~\ref{fig_amq_plaq_nf2_quad} we show the results for the plaquette.
For each $\beta$ value the data appear to be rather
linear in the quark mass $am_q$, in particularly for the higher
$\beta$ values. This suggests that a quadratic fit ansatz is sufficient
to obtain the value of $P$ in the chiral limit. We thus use
\begin{equation}
   P = d_0 + d_1 am_q + d_2 (am_q)^2 \,.
\label{plaq_fit_eq}
\end{equation}
Except for $\beta = 5.20$, it does not much matter whether a linear
or quadratic fit is used.
In Fig.~\ref{fig_amq_r0oa_nf2_joint8para_15pts_041221} we show the results for
$r_0/a$. The data is less linear in $am_q$, and also less smooth,
so we used the renormalisation group inspired global fit ansatz
\begin{equation}
   \ln {r_0\over a}
       =  A_1(\beta) + A_2(\beta) am_q + A_3(\beta) (am_q)^2  \,,
\end{equation}
where $A_1(\beta)$ is a linear polynomial in $\beta$,
and $A_2(\beta)$, $A_3(\beta)$ are quadratic polynomials in $\beta$.
This ansatz was also used in \cite{gockeler04a}. The results of the fits in
the chiral limit are given in Table~\ref{table_critical}.

\begin{figure}[!t]
\begin{center}
   \epsfxsize=10.00cm 
      \epsfbox{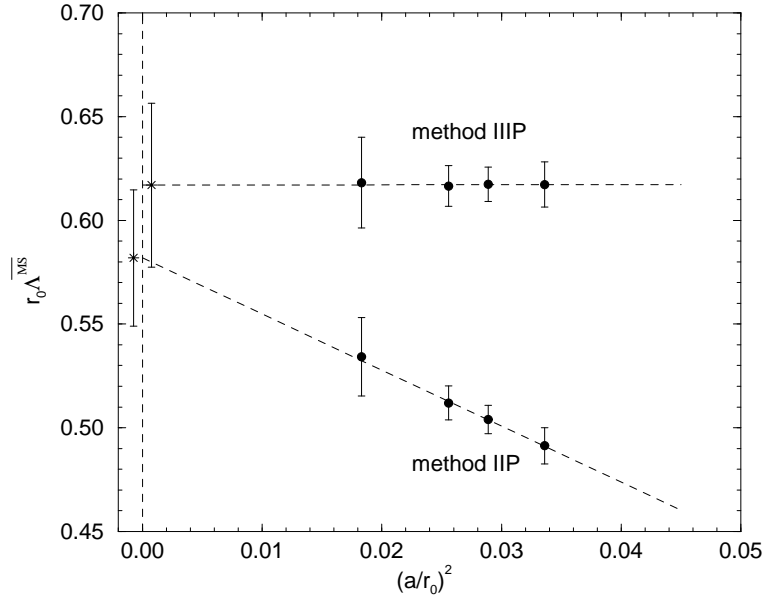}
\end{center}
   \caption{The unquenched $r_0\Lambda^{\msbar}$ points (filled circles)
            versus $(a/r_0)^2$, together with a linear extrapolation
            to the continuum limit for methods IIP and
            IIIP. Stars represent the extrapolated values.}
   \label{fig_r0lamMSbar}
\end{figure}

In Table~\ref{table_r0_lam_dyn} we give our results for $r_0\Lambda^{\msbar}$
for methods I, II, IIP, III and IIIP. Again, as the results for method I are
very similar to method II, we shall not discuss method I further here.
In Fig.~\ref{fig_r0lamMSbar} we plot $r_0\Lambda^{\msbar}$ against $(a/r_0)^2$
for methods IIP and IIIP, together with a linear extrapolation to the
continuum limit. Though we cannot reach such small $a$ values as for the
quenched case, the $r_0\Lambda^{\msbar}$ data do seem to lie
on straight lines. We find a linear behaviour at least over the region $a^{-1}
\sim 2$ -- $3\,\mbox{GeV}$. This seems to be well inside the linear region of 
Fig.~\ref{fig_r0lamMSbar_lbfun_nf0_IIP_8pts}.

For methods IIP (and II) the results lie roughly parallel to the quenched
results, while for methods IIIP (and III) they are flatter and higher.
However, in the continuum limit they agree within error bars. 
Ideally, the result should not depend on the choice of trajectory.
The way this should work, as mentioned before, is that although
the coefficients $t_i^{\plaq}$ will be different depending on the
path one might choose, the sum
\begin{equation}
\frac{1}{g_{\plaq}^2(a)} - t_1^{\plaq} -t_2^{\plaq}\,g_{\plaq}^2(a) + \ldots
\end{equation}
should not. However, at the order to which we have the series this is
not yet so. The difference between methods II and III is that we have
replaced $c_{sw}^{\plaq}$ by its 1-loop expansion. 
Returning to Fig.~\ref{fig_r0lamMSbar}, the fact that the
results from methods II, IIP are almost parallel to the quenched results 
suggests that in methods II, IIP the $O(a^2)$ effects come from the
same source as in the quenched case, which must be the gluon action. For
methods III, IIIP the slope is much smaller so there must have been a
fortuitous cancellation between $a^2$ effects from the gluon and
fermion terms.

One expects that Pad{\'e} improvement gives a better answer, so the 
P results are more trustworthy. Previous experience suggests that 
the procedure in IIP of using tadpole improved $c_{sw}$ works fairly well. For 
example, $\kappa_c$ in \cite{gockeler97a} and the renormalisation constant
$Z$ for $v_{2b}$ in \cite{gockeler04c} agree within a few percent
with the non-perturbative values. However, method IIIP is a more consistent
approach. 
Furthermore, the results from method IIIP appear to be insensitive to the
particular form of the continuum extrapolation. We therefore take these
numbers as our best estimate.

From the linear extrapolation of method IIIP to the continuum limit we thus
quote 
\begin{equation}
  r_0\Lambda^{\msbar}_2 \equiv \left. r_0\Lambda^{\msbar}\right|_{n_f=2} =
  0.617(40)(21) \,, 
\label{res2}
\end{equation}
where the first error is statistical and the second systematic. The latter
error is estimated by the spread in the results between method III and IIIP.
Compared to our previous result~\cite{booth01a}, the value (\ref{res2}) has
increased by $\approx 10\%$, but still lies within the error bars.


\section{Extrapolation to $n_f=3$ Flavours}
\label{force_scales}


At high energy scales we can see that $\Lambda^{\msbar}$ makes some 
fairly large jumps as we pass through the heavy quark mass 
thresholds and change the effective number of flavours. 
From \cite{chetyrkin97a} we can see that the reason for these 
large jumps is the fact that $m_q /\Lambda^{\msbar}$ is large.
We want to argue here that the situation with light quarks,
$m_q \lesssim \Lambda^{\msbar}$, is rather different, and that in
this case we do not expect to see any dramatic dependence of
$\Lambda^{\msbar}$ on $n_f$. 

We will determine the $n_f=3$ flavour $\Lambda$ parameter from matching the
static force at the scale $r_0$. 


 \subsection{One-loop Matching} 


To make clear what is involved in matching, we will go through 
the 1-loop calculation in some detail. 

At the 1-loop level the static potential between fundamental
charges is given by 
\begin{equation}
\begin{split}
    V(r) = - \frac{4}{3} \frac{g^2_{\msbar}(\mu)}{4 \pi r} 
       \Bigg\{ 1 + \frac{g^2_{\msbar}(\mu)}{16 \pi^2}   
       \Bigg[ &22 \left( \ln \mu r + \gamma_E + \frac{31}{66}  \right)  \\
        &- \frac{4}{3} n_f 
          \left( \ln \mu r + \gamma_E +\frac{5}{6} \right)
          \Bigg] +\cdots \Bigg\}  
\end{split}
\end{equation} 
for massless sea quarks (see, for example,~\cite{MM}). 
We can work out the force $f(r)$ at distance $r$ by differentiating this 
to give 
\begin{equation}
\begin{split}
    4 \pi r^2 f(r) = \frac{4}{3} g^2_{\msbar}(\mu)
     \Bigg\{ 1 + \frac{g^2_{\msbar}(\mu)}{16 \pi^2} \Bigg[ &22 
        \left( \ln \mu r +  \gamma_E -\; \frac{35}{66} \right)    \\
      &-  \frac{4}{3} n_f
         \left( \ln \mu r + \gamma_E -\;\frac{1}{6} \right)
         \Bigg] +\cdots \Bigg\} \,. 
\end{split}
\label{f1loop}
\end{equation}
If we now change the flavour number from 2 to 0, or from 2 to 3,
while keeping the force at distance $r$ constant, we get
\begin{equation} 
\begin{split}
 33 \ln \frac{\Lambda_0^{\msbar}}{\Lambda_2^{\msbar}} &=
  -4 \left(\ln \Lambda_2^{\msbar} r + \gamma_E - \frac{1}{6} \right) \,,  \\
 (33 -6) \ln \frac{\Lambda_3^{\msbar}}{\Lambda_2^{\msbar}} &= 
   2 \left(\ln \Lambda_2^{\msbar} r + \gamma_E - \frac{1}{6} \right)  \,.
\end{split} 
\end{equation} 
 We can eliminate $r$ from these equations, leaving us with the 
 simple equation
 \begin{equation} 
    \frac{\Lambda_3^{\msbar}}{\Lambda_2^{\msbar}}
       = \left(\frac{\Lambda_2^{\msbar}}{\Lambda_0^{\msbar}}
             \right)^{\frac{11}{18}} \,,
 \end{equation} 
 which can be used to estimate $\Lambda_3^{\msbar}$ from the $n_f=0$ and
 $n_f=2$ results.


\subsection{Higher Loops} 


To repeat this matching calculation with more loops, we follow
\cite{necco01a} and define a force-scale coupling $g_{\qq}$ by
\begin{equation}
   4 \pi r^2 f(r) \equiv \frac{4}{3} g^2_{\qq}(r) \,.
\end{equation} 
From eq.~(\ref{f1loop}) we can read off
\begin{equation}
   t_1^{\qq} = -\; \frac{1}{(4 \pi)^2} 
   \left[ 22 \left( \gamma_E -\; \frac{35}{66} \right)
   -\frac{4}{3} n_f  \left( \gamma_E -\; \frac{1}{6} \right) \right] \,.
\label{t1qq}
\end{equation} 
We can find $t_2^{\qq}$ by calculating the force from the 2-loop
expression of $V(r)$ reported in~\cite{schroder98a,peter97a}:
\begin{equation}
\begin{split}
 t_2^{\qq} &=  \frac{1}{(4 \pi)^4} \Bigg[ 
 \frac{1107}{2} - 204 \, \gamma_E - \; \frac{229}{3} \pi^2 
 + \frac{9}{4} \pi^4  -66\, \zeta_3 \\
 & \hspace*{2cm} 
 + \frac{n_f}{3} \left( - \; \frac{553}{3} + 76 \gamma_E +
 \frac{44}{3} \pi^2 + 52\, \zeta_3 \right) 
 + \frac{4}{27} n_f^2\, ( 12 -\pi^2 ) \Bigg] \,,
\end{split}
\end{equation} 
which gives us enough information to calculate the 3-loop $\beta$ function
for $g_{\qq}(r)$ (cf eq.~(\ref{bi_lat})).
There would be complications in going to the next order, 
because it is known that terms of the type $\alpha_s^4 \ln \alpha_s$ 
will enter the series for the potential~\cite{appelquist78a}.
 
\begin{figure}[!t]
   \begin{center}
   \epsfig{file=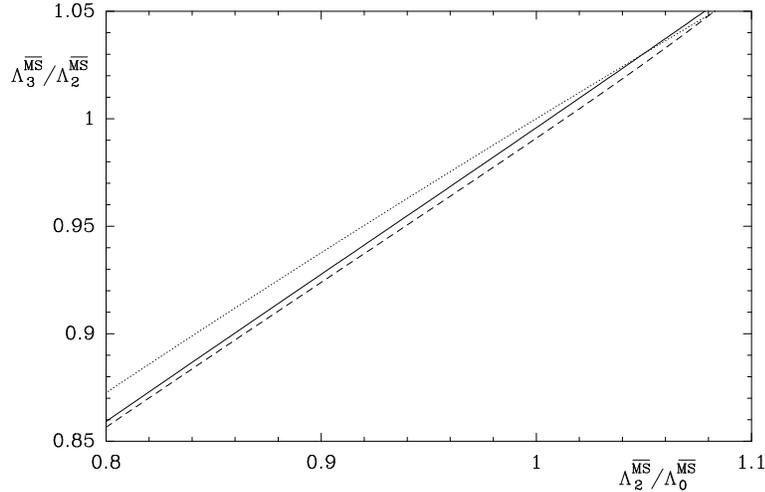,angle=270,width=10cm}
   \end{center}
   \caption{The ratio $\Lambda^{\msbar}_3/\Lambda^{\msbar}_2$ against the
     ratio 
     $\Lambda^{\msbar}_2/\Lambda^{\msbar}_0$ from 1-loop (dotted line),
     2-loop (dashed line) and 3-loop (solid line) matching.}
   \label{fig_Splot}
\end{figure}

We are now ready to see how $\Lambda^{\msbar}$ depends on flavour number, if
we make the value of $f(r)$ independent of $n_f$ (the number of massless
quark flavours) at some particular $r$ value. Implicitly, we assume $r
\approx r_0$. 
If $f(r)$ is independent of $n_f$, then $g_{\qq}(r)$ is independent of $n_f$
too.  We can compare the $q\bar{q}$ scheme $\Lambda$s by using
\begin{equation}
\begin{split}
   r \Lambda^{\qq}_{0} &= F^{\qq}(g_{\qq}(r), n_f=0) \,, \\
   r \Lambda^{\qq}_{2} &= F^{\qq}(g_{\qq}(r), n_f=2) \,, \\
   r \Lambda^{\qq}_{3} &= F^{\qq}(g_{\qq}(r), n_f=3) \,. 
\end{split}
\label{Feq}
\end{equation} 
We can take ratios of these equations to cancel $r$
and find equations for $\Lambda$ ratios. 
These  $q\bar{q}$ scheme $\Lambda$ ratios can then be converted
into $\overline{MS}$ by using $t_1^{q \bar{q}}$ from eq.~(\ref{t1qq}).
This gives us a way of making a parametric plot
of $\Lambda$ ratios by varying $g_{\qq}$ and calculating all three
$\Lambda$s from $g_{\qq}$. In Fig.~\ref{fig_Splot} we show the plot. 

The results clearly have to be treated with some caution, because
$r_0 \Lambda$  is a fairly large number. So it is not clear how much
we can learn from perturbative results at the scale $r_0$.
It is therefore quite surprising that the 
different orders of perturbation theory agree so well in Fig.~\ref{fig_Splot}.
Furthermore, we have assumed in this section that $r_0 m_s \ll 1$, so that
the strange quark can reasonably be treated as massless. Both these
difficulties could be decreased by using a smaller distance (and thus a 
smaller value for $r^2 f(r)$) to set our scale. 

\subsection{Result for $n_f=3$} 

From our quenched and unquenched $n_f=2$ results, (\ref{res0}) and
(\ref{res2}), we obtain $\Lambda_2^{\msbar}/\Lambda_0^{\msbar} =
1.005$. If we insert this number into the 3-loop matching curve shown in
Fig.~\ref{fig_Splot}, we find $\Lambda_3^{\msbar}/\Lambda_2^{\msbar} =
0.999$. From this ratio and eq.~(\ref{res2}) we then obtain for
$n_f = 3$ quark flavours 
\begin{equation}
  r_0\Lambda^{\msbar}_3 \equiv \left. r_0\Lambda^{\msbar}\right|_{n_f=3} =
  0.616(29)(19) \,. 
\end{equation}
We have not attempted to estimate the systematic error induced by the matching
procedure.


\section{Comparison with Phenomenology}
\label{phenomenology}


In this section we shall make a comparison with other lattice
and phenomenological results. For this we first need to set the force
scale in terms of a physical unit. 

To fix the scale $r_0$ in physical units, we extrapolate recent dimensionless
nucleon masses $m_N r_0$ found by the CP-PACS, JLQCD and QCDSF-UKQCD
collaborations jointly to the physical pion mass following~\cite{alikhan04a}.
This gives the value $r_0=0.467\,\mbox{fm}$ with an estimated error of
$7\%$. We will use this number throughout this paper. A similar result for
$r_0$ was quoted in~\cite{aubin04a}.  

For the quenched case we then obtain
\begin{equation}
   \Lambda^{\msbar}_0 = 259(1)(20)\,\mbox{MeV} \,,
\end{equation}
and for the unquenched case we find
\begin{eqnarray}
   \Lambda^{\msbar}_2 &=& 261(17)(26)\,\mbox{MeV} \,, \\
   \Lambda^{\msbar}_3 &=& 260(12)(26)\,\mbox{MeV} \,.
\end{eqnarray}
The systematic errors quoted here include the uncertainty in setting the
scale. Note that previously~\cite{booth01a} we had assumed $r_0 =
0.5\,\mbox{fm}$.    

\begin{figure}[t]
\begin{center}
   \epsfxsize=10.00cm \epsfbox{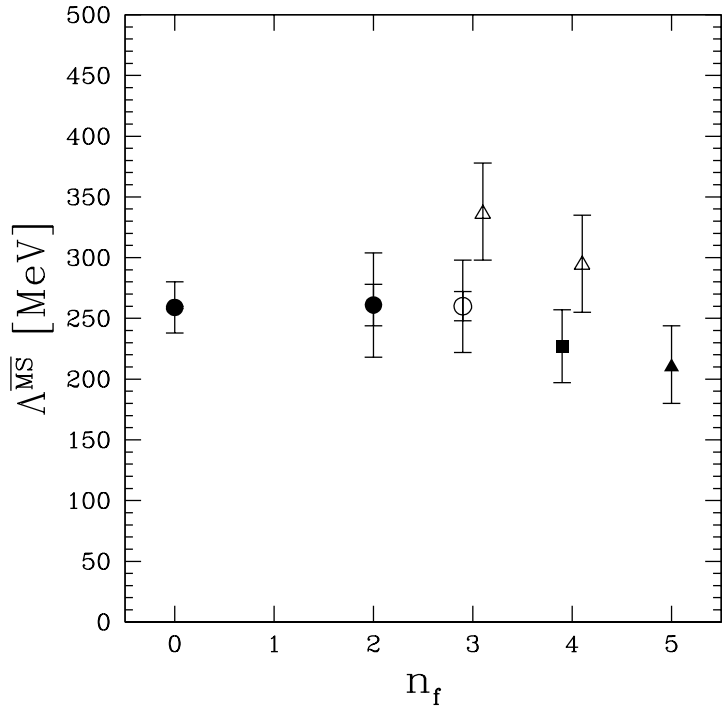}
\end{center}
\caption{Values of $\Lambda^{\msbar}$ versus number of quark flavours
$n_f$. The filled circles are our $n_f = 0$, $2$ results, and the open
circle is our extrapolated value. The inner error bars give the statistical
errors, while the outer error bars give the total errors. The square is from a
3-loop analysis of the non-singlet structure functions~\cite{blumlein04a}. The
triangles are taken from~\cite{bethke02/04}. The open triangles are evaluated
using the 4-loop expansion of $\alpha_s$ and 3-loop matching at the quark
thresholds. The entries at $n_f=3$ and $4$ have been displaced horizontally.}
\label{fig_lambda_nf}
\end{figure}

In Fig.~\ref{fig_lambda_nf} we show our results for $\Lambda^{\msbar}$
together with recent experimental values from \cite{blumlein04a} and
\cite{bethke02/04}. It appears that the lattice results extrapolate smoothly
to the experimental values at $n_f=4$~\cite{blumlein04a} and
$n_f=5$~\cite{bethke02/04}. However, our $n_f=3$ result 
lies two standard deviations below the corresponding phenomenological value
(open triangle). (The reader should be aware that the sometimes called
experimental numbers imply a good deal of modelling and, thus, should be
regarded as phenomenological numbers.) 

\begin{figure}[!t]
\begin{center}
   \epsfxsize=10.00cm \epsfbox{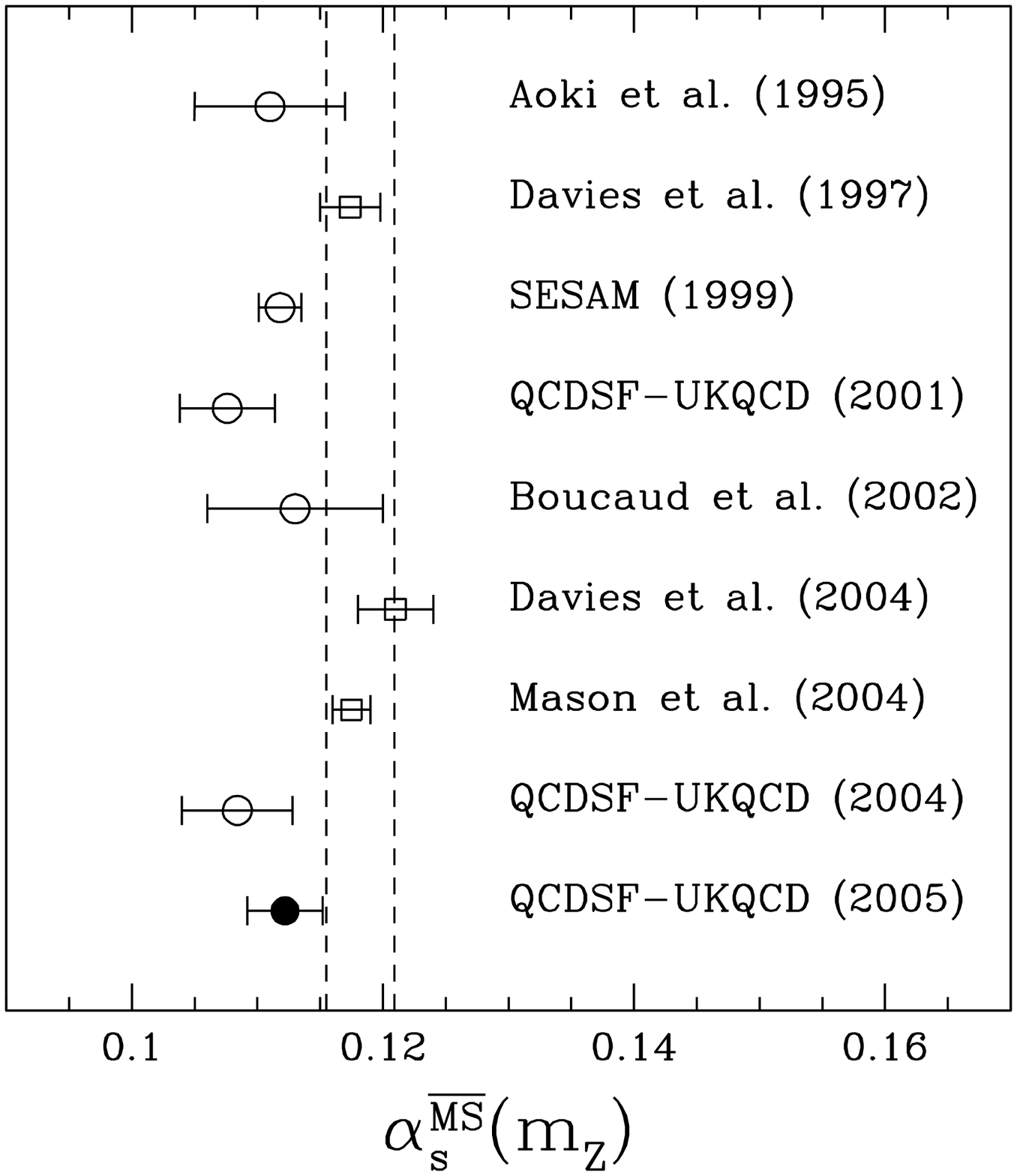}
\end{center}
\caption{Comparison of $\alpha_s^{\msbar}(m_Z)$ from this work (solid circle)
with other lattice results~\cite{aoki95a,davies97a,sesam99a,booth01a,%
boucaud02a,davies03,mason04a,gockeler04b} (from top to bottom). The circles
are from  
Wilson fermions and the squares from staggered fermions. The dashed line
indicates the mean experimental value~\cite{bethke02/04}.}
\label{fig_alphalat}
\end{figure}

In order to compare $\alpha_s$ from various experiments and theory, it must be
evolved to a common scale. For convenience this is taken to be the mass of the
$Z$ boson, $m_Z$. Having computed $\Lambda^{\msbar}$ for $n_f=3$ flavours, we
may use the 4-loop expansion of $\alpha_s$ and the 3-loop matching condition
at the quark thresholds~\cite{chetyrkin97a,chetyrkin98a} to determine
$\alpha^{\msbar}_{n_f=5}(m_Z)$. We take the charm and bottom thresholds to be
at $1.5$ and $4.5\,\mbox{GeV}$, respectively. Furthermore, we choose the charm
and bottom quark masses to be $m_c^{\msbar}(m_c) = 1.5\,\mbox{GeV}$ and
$m_b^{\msbar}(m_b) = 4.5\,\mbox{GeV}$, respectively. Varying the charm and
bottom quark masses within reasonable limits has a neglible effect on the
final result. We then obtain 
\begin{equation}
\alpha_{n_f=5}^{\msbar}(m_Z) = 0.112(1)(2)\,.
\label{alphasmZ}
\end{equation}
This is to be compared with the world average value~\cite{bethke02/04}
$\alpha_s^{\msbar}(m_Z) = 0.1182(27)$.

In Fig.~\ref{fig_alphalat} we compare our result for $\alpha_s^{\msbar}(m_Z)$
with other lattice results and experiment. We find agreement with previous 
lattice calculations using Wilson fermions. It occurs that the Wilson  
results lie systematically below the mean experimental value. On the other
hand, calculations using staggered fermions (albeit from the same group) show
a better agreement with experiment. Our result for $r_0 \Lambda^{\msbar}_2$
agrees 
also with that of the ALPHA collaboration~\cite{DellaMorte04a}, which does not
quote a number for $\alpha_s^{\msbar}(m_Z)$. Our result for
$\alpha_s^{\msbar}(m_Z)$ lies two standard deviations below the
phenomenological value.   



\section{Conclusions}
\label{conclusions}

Due to substantial improvements of the performance of our hybrid Monte Carlo
algorithm~\cite{Bakeyev03a}, we were able to extend our dynamical simulations
to smaller quark masses and to larger values of $\beta$. Our smallest lattice
spacing now is $a \approx 0.07\,\mbox{fm}$. This enabled us to perform a
chiral and 
continuum extrapolation of the lattice data. Because the calculation involves
a perturbative conversion from the lattice coupling constant to the
(mass independent) $\overline{MS}$ 
constant, it was important to first extrapolate the lattice data to the chiral
limit. We have discussed basically two approaches of converting the lattice
coupling constant to the $\overline{MS}$ one. They differed mainly in how the
non-perturbative improvement (clover) term was incorporated in the
perturbative expansion. It was reassuring to see that both methods led to the
same result in the continuum limit. This indicates once more that a reliable
extrapolation to the continuum limit is very important.

We could also improve on our quenched result, because data at smaller lattice
spacings became available. 

There are several sources of systematic error in our calculation. The main 
error comes from setting the scale, followed by the continuum extrapolation. 
As better dynamical data become available, the uncertainty in setting the
scale will be gradually reduced. Simulations at smaller lattice spacings will
become possible with the next generation of computers, which should facilitate
the extrapolation to the continuum limit.         


\section*{Acknowledgements}

We like to thank Antonios Athenodorou and Haris Panagopoulos for checking the
numbers in eq.~(\ref{check}).
The numerical calculations have been performed on the Hitachi SR8000 at
LRZ (Munich), on the Cray T3E at EPCC (Edinburgh)
\cite{allton01a},
on the Cray T3E at NIC (J\"ulich) and ZIB (Berlin),
as well as on the APE1000 and Quadrics at DESY (Zeuthen).
We thank all institutions.
This work has been supported in part by
the EU Integrated Infrastructure Initiative Hadron Physics (I3HP) under
contract RII3-CT-2004-506078
and by the DFG under contract FOR 465 (Forschergruppe
Gitter-Hadronen-Ph\"anomenologie).



\end{document}